\documentclass[aps,prl,twocolumn,noshowpacs,superscriptaddress]{revtex4-1}
\usepackage{graphicx,epsfig, subfigure}
\usepackage{amsbsy}
\usepackage{amssymb}
\usepackage{amsmath}
\usepackage{hyperref}
\usepackage[bottom]{footmisc}
\usepackage{color}

\date{\today}

\begin{document}

\newcommand{\eqnref}[1]{Eq.~\ref{#1}}
\newcommand{\figref}[2][]{Fig.~\ref{#2}#1}

\title{
Electron spin control of optically levitated nanodiamonds in vacuum
}
\author{Thai M. Hoang}
 \affiliation{Department of Physics and Astronomy, Purdue University, West Lafayette, IN 47907, USA}

\author{Jonghoon Ahn}
 \affiliation{School of Electrical and Computer Engineering, Purdue University, West Lafayette, IN 47907, USA}

\author{Jaehoon Bang}
 \affiliation{School of Electrical and Computer Engineering, Purdue University, West Lafayette, IN 47907, USA}

\author{Tongcang Li}
 \email{Correspondence and requests for materials should be addressed to T.L. (email: tcli@purdue.edu).}
 \affiliation{Department of Physics and Astronomy, Purdue University, West Lafayette, IN 47907, USA}
 \affiliation{School of Electrical and Computer Engineering, Purdue University, West Lafayette, IN 47907, USA}
 \affiliation{Birck Nanotechnology Center, Purdue University, West Lafayette, IN 47907, USA}
 \affiliation{Purdue Quantum Center, Purdue University, West Lafayette, IN 47907, USA}

\begin{abstract}
Electron spins of diamond nitrogen-vacancy (NV) centers are important quantum resources for nanoscale sensing and quantum information. Combining NV spins with levitated optomechanical resonators will provide a hybrid quantum system for  novel applications. Here we optically levitate a nanodiamond  and demonstrate electron spin control of its built-in NV centers in low vacuum. We observe that the strength of electron spin resonance (ESR) is enhanced when the air pressure is reduced. To better understand this system, we  investigate the effects of trap power and measure the absolute internal temperature of  levitated nanodiamonds with ESR after calibration of the  strain effect. We also observe that oxygen and helium gases have different effects on both the photoluminescence and the ESR contrast of nanodiamond NV centers, indicating  potential applications of NV centers in  oxygen gas sensing. Our results  pave the way towards a levitated  spin-optomechanical system for studying macroscopic quantum mechanics.
\end{abstract}
\maketitle

\section{Introduction}
With electron spin states accessible by visible lasers and microwave radiation \cite{Clark1971, Loubser1978, Gruber1997}, diamond nitrogen-vacancy (NV) centers have broad applications in quantum information \cite{Neumann2008, Togan2010, neumann2010quantum} and nanoscale sensing of  magnetic field \cite{Degen2008, Balasubramanian2008, Taylor2008}, temperature \cite{Acosta2010,Chen2011}, and beyond \cite{Steinert2013,Schirhagl2014}.
Combining such NV spin systems with mechanical resonators \cite{kolkowitz2012coherent,arcizet2011single} will provide a hybrid system with versatile functions \cite{yin2015review}.
Recently, several authors  proposed to levitate a nanodiamond with a built-in NV center in vacuum as a novel hybrid spin-optomechanical  system  for creating large quantum superposition states \cite{Yin2013, Scala2013}, testing quantum wavefunction collapse models \cite{RomeroIsart2011, Yin2013, Scala2013} and quantum gravity \cite{albrecht2014testing}. A levitated nanodiamond will also be an ultrasensitive torque detector \cite{Hoang2016Torsional} and mass spectrometer at room temperature \cite{Zhao2014}. These proposals are inspired by recent progresses in  optomechanics of levitated pure dielectric (mainly silica) particles \cite{romero2010toward,Chang2010,Li2010,Geraci2010,Li2011,Gieseler2012,Arvanitaki2013,yin2013review}, which show optical levitation in vacuum can isolate the mechanical motion of a dielectric particle from the environment to improve the mechanical quality factor. 

\begin{figure}[h!]
	\includegraphics[scale=0.99]{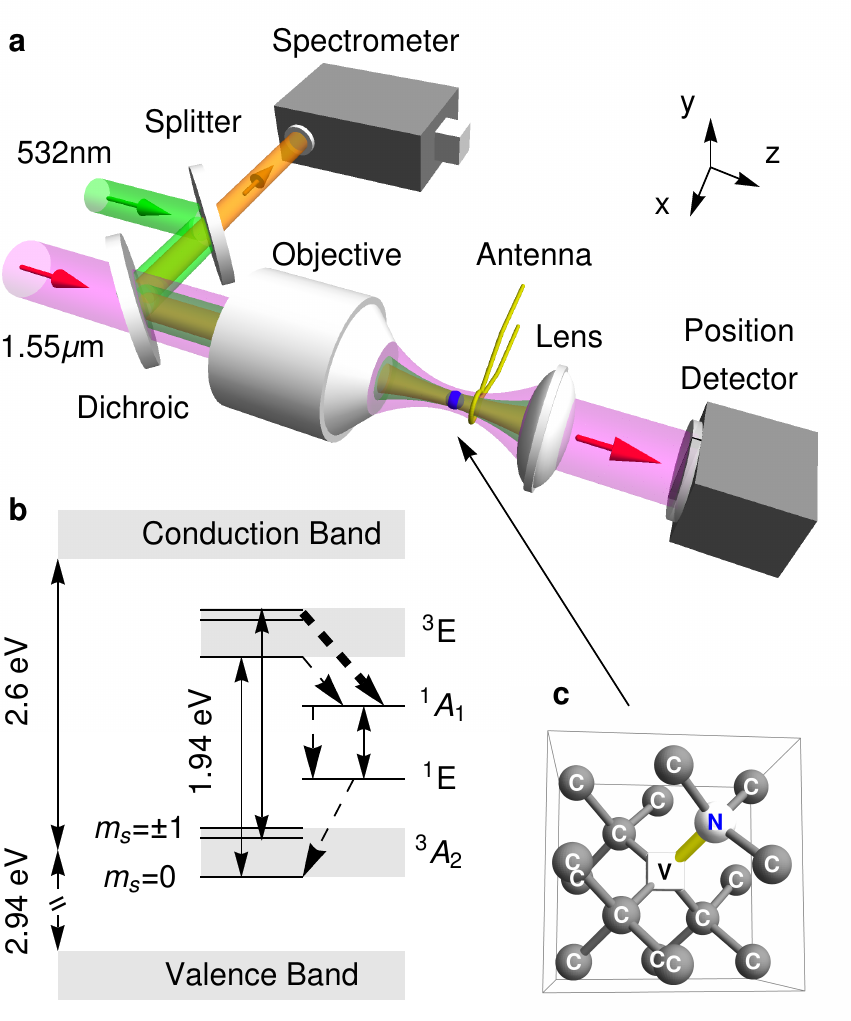}
	\caption{ \textbf{Schematic of the experiment.} \textbf{a} A nanodiamond (magenta sphere) is trapped inside the vacuum chamber using optical tweezers formed by a 1550 nm laser (magenta beam)  and an objective lens. The position detector monitors its center-of-mass motion. The NV centers are optically excited by a 532 nm laser (green beam) guided by a beam splitter and a long-pass dichroic mirror. The fluorescent signal (orange beam) of the NV centers is detected by a spectrometer with an EMCCD camera. Electron spins are controlled by  microwaves delivered by an antenna.  Magenta, green, and orange arrows illustrate the directions of light beams. \textbf{b}, Energy levels of a NV$^-$ center. Solid arrows represent radiative transitions while  dashed arrows represent nonradiative transitions.  $^3A_2$, $^1E$, $^1A_1$, and $^3E$ denote electronic states of the NV center, and $m_\mathrm{s}=0,\pm 1$ denote the spin states of $^3A_2$. \textbf{c}, Molecular structure of a NV center in a nanodiamond (the blue sphere in \textbf{a} indicated by the arrow). Carbon atoms are shown in gray spheres (labeled C), the lattice vacancy is labeled by a white cube (labeled V) and the nitrogen atom is labeled by a white sphere (labeled N). Each nanodiamond has about 500 NV centers.}
	\label{Fig:Apparatus}
\end{figure}

Nanodiamonds have been optically trapped in liquid \cite{Horowitz2012, Geiselmann2013three}, atmospheric air \cite{Neukirch2013} and low vacuum \cite{Neukirch2015, Rahman2016, Hoang2016Torsional}. Recently, the electron spin resonance of optically levitated nanodiamonds  has been demonstrated in atmospheric air \cite{Neukirch2015}.  The angular trapping and  torsional vibration of an optically levitated nanodiamond have also been observed, which are important for controlling the orientations of NV spins \cite{Hoang2016Torsional}.  The next key step towards realizing these ambitious proposals \cite{Yin2013, Scala2013, albrecht2014testing, Zhao2014} is to control the NV electron spin of a levitated nanodiamond in a vacuum environment, which has not been demonstrated before.

Here we demonstrate the electron spin control and direct temperature measurement of  NV centers in an optically levitated nanodiamond  in low vacuum. Our work is enabled by using a 1550 nm trapping laser. We find that the 1550 nm laser is more benign to the photoluminescence of NV centers than a 1064 nm trapping laser \cite{Neukirch2015}. One would usually expect that the ESR signal of an optically levitated nanodiamond in vacuum to be  weaker  than that in atmospheric air due to laser heating in vacuum \cite{Toyli2012}. To our surprise, the electron spin resonance (ESR) contrast of an optically levitated nanodiamond is  enhanced in a vacuum environment. We attribute this ESR enhancement to the reduction of low-quality negatively-charged NV$^-$ centers near the surface due to the reduction of oxygen surface termination \cite{FuAPL2010, HaufPRB2011, Grotz2012, OhashiNL2013}, and a moderate increase of the temperature that quenches low-quality surface NV$^-$ centers \cite{Plakhotnik2010} without significantly affecting high-quality NV$^-$ centers at the center of the nanodiamond\cite{Toyli2012}. We also observe that oxygen (O$_2$) and helium (He) gases have  different effects on the photoluminescence and ESR contrast of nanodiamond NV centers. These observed effects are reversible, indicating that nanodiamond NV centers can be used for  oxygen gas sensing \cite{Schirhagl2014}. To better understand this system, we investigate the effects of trapping laser on ESR and measure the absolute temperature of levitated nanodiamonds in vacuum with ESR \cite{Acosta2010,Toyli2012} after calibration of the strain effect \cite{Schirhagl2014}. The ability to remotely  measure the internal temperature of a levitated nanodiamond will have applications in investigating nonequilibrium thermodynamics \cite{Li2010, Millen2014, Gieseler2014}.

\section{Results}
\textbf{Optical levitation of nanodiamonds.}
In our experiment, a nanodiamond with a diameter about 100 nm is optically trapped inside a vacuum chamber (\figref{Fig:Apparatus}). We first launch  a mixture of commercial nanodiamonds and water to the trapping region with an ultrasonic nebulizer. The water microdroplets evaporate and a nanodiamond will be  captured  by a tightly focused 500~mW 1550~nm laser beam after waiting for a few minutes (see Methods). After a nanodiamond is captured, we  evacuate the air in the  chamber by a turbomolecular pump. A trapped nanodiamond will eventually be lost when the air pressure is below a critical value, which is a few Torr for commercial nanodiamonds  that we used (see Methods).

\begin{figure}[t]
\begin{minipage}{3.5in}
\includegraphics[scale=0.9]{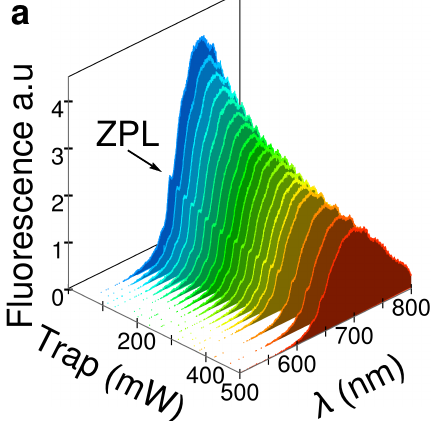} \\
\includegraphics[scale=0.9]{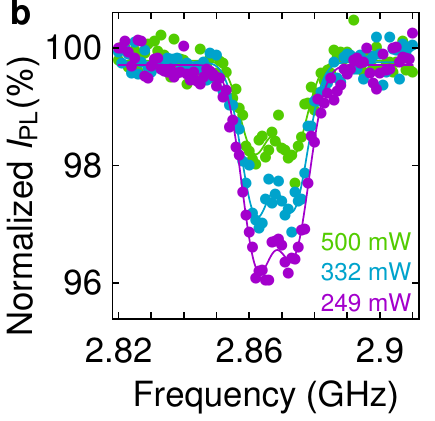}
\includegraphics[scale=0.9]{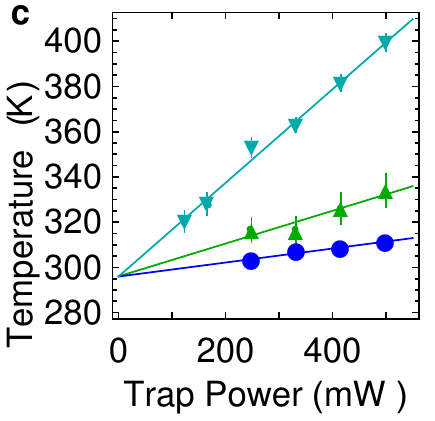}
\end{minipage}
\caption{\textbf{Effects of optical trapping power.} \textbf{a}, The  fluorescence strength increases by about 2.5 times when the power of the trapping laser is reduced from 500~mW to 60~mW. The zero phonon line (ZPL) is indicated by the arrow. The nanodiamond is excited with a 260 $\mu W$ green laser. \textbf{b}, Electron spin resonance (ESR) spectra of NV$^-$ centers at different trapping powers. Normalized $I_\mathrm{PL}$ is the ratio of the total fluorescence count with and without microwave excitation. The ESR spectra data  are taken with a green laser of 30~$\mu W$ and fitted with the double Gaussian function. Each ESR scan takes about 30 minutes to achieve a high signal-to-noise ratio. These ESR spectra are taken with the nanodiamond marked as blue circle in \textbf{c}, which has a small temperature change.
\textbf{c}, At the initial trapping power of 500~mW, the internal temperature of the nanodiamond ranges between 300-400~K depending on  individual nanodiamond. When the trapping power reduces, the internal temperature of the nanodiamond approaches the room temperature (296~K). Temperatures are extracted from the double Gaussian fits of the ESR spectra, and the error bars of temperatures are obtained from the standard errors of the fitted parameters from the ESR spectra. Each marker shape represents a different nanodiamond. Data are acquired at atmospheric pressure. a.u. denotes an arbitrary unit.}
\label{Fig:NVpower}
\end{figure}

\textbf{Effects of trapping laser on electron spin resonance.}
For an optically levitated nanodiamond, one needs to consider  effects of the trapping laser  while studying its NV centers. We investigate this effect by measuring the fluorescence spectra and ESR spectra at different trapping powers (\figref{Fig:NVpower}). We excite the trapped nanodiamond with a 532~nm laser and collect its fluorescent signal by a spectrometer with an EMCCD (electron multiplying charge coupled device) camera with a single-photon sensitivity. To study the ESR, we excite the electron spin states  by microwave radiation delivered by a coplanar waveguide antenna (see Methods). The normalized fluorescence signal $I_\mathrm{PL}$ of the ESR spectrum is the ratio of total fluorescence count with and without the microwave excitation. The resonance peaks of electron spin transition occur at the microwave frequencies of minimum $I_\mathrm{PL}$. The ESR contrast is defined as $1-I_\mathrm{PL}$. As shown in  \figref{Fig:NVpower}{a,b}, both the fluorescent signal and the ESR contrast decrease when the trapping power increases.
To understand these  phenomena, we need to consider two primary effects associated with the trapping laser:  heating \cite{Lai2013} and photo-induced ionization when a 532 nm laser is also present \cite{Aslam2013}.

\begin{figure*}[t]
\begin{minipage}{7in}
	\includegraphics[scale=0.9]{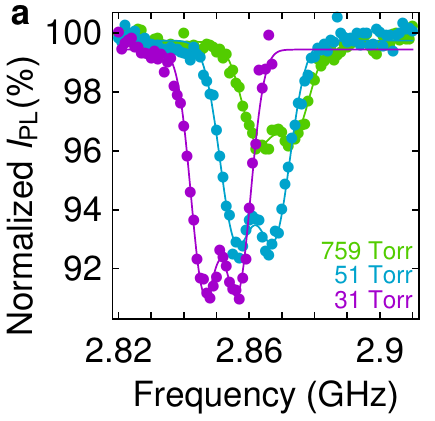}	
\includegraphics[scale=0.9]{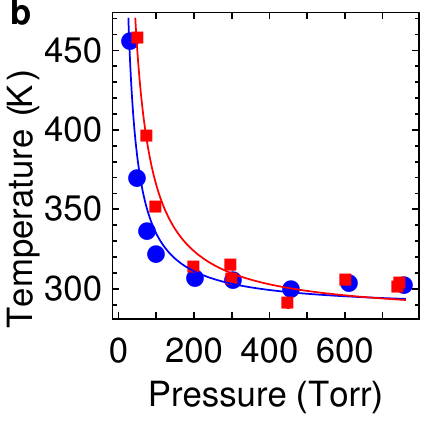}
\includegraphics[scale=0.9]{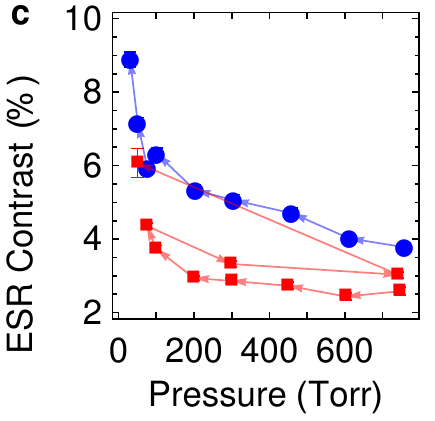}	
\end{minipage}
\caption{\textbf{Electron spin resonance in low vacuum.}
\textbf{a}, ESR spectra of a levitated nanodiamond at different air pressures. Each ESR scan takes approximately 30 minutes to achieve a high signal-to-noise ratio. The peaks of the ESR spectra shift in vacuum due to the thermal effect. The ESR spectra data are fitted with double Gaussian functions.
\textbf{b}, The measured temperature of two different nanodiamonds as a function of air pressure. From atmospheric pressure to low vacuum, their temperatures change from 300~K to 450~K and beyond. The errorbar of temperature measurement is smaller than the marker size. Each marker shape represents a different nanodiamond.
\textbf{c}, The maximum contrast of each ESR spectrum increases as the air pressure decreases. Arrows of the lines indicate the time order of the experiment.
The trapping power for each nanodiamond is always held constant.  a.u. denotes an arbitrary unit. 
The error bars of temperature measurements and ESR contrast are obtained from the standard errors of the fitted parameters of the ESR spectra.
}
\label{Fig:NVpressure}
\end{figure*}

Since the  absorption of the trapping laser increases  temperature, a higher trapping power leads to a higher temperature of the nanodiamond (\figref{Fig:NVpower}{c}).
The visible fluorescent signal comes from the radiative decay transition $^3E$  $\rightarrow$ $^3A_2$. There is also a competing process that  the excited state $^3E$ may relax to the ground state $^3A_2$ through nonradiative process with metastable states $^1E$ and $^1A_1$  (\figref{Fig:Apparatus}{b}).
Because a high temperature enhances the nonradiative decay \cite{Toyli2012}, it weakens the visible fluorescence strength (\figref{Fig:NVpower}{a}).

Besides heating, the 1550 nm trapping laser can  cause photo-induced ionization of NV$^-$ to NV$^0$  when the 532 nm laser is on  \cite{Aslam2013, Iakoubovskii2000, Kupriyanov2000, Manson2005}.
As shown in \figref{Fig:Apparatus}{b}, the 532 nm laser excites the NV centers from the ground state $^3A_2$ to the excited state $^3E$. Since the separation between the excited state and the conduction band is 0.67~eV (corresponding to a vacuum wavelength of 1880~nm) \cite{Aslam2013}, the 1550 nm trapping beam can excite electrons from the excited state $^3E$ to the conduction band and consequently ionize NV$^-$ to NV$^0$.
Because high trapping power increases the ionization rate from NV$^-$ to NV$^0$, the fluorescence signal from NV$^-$ will be reduced as seen in \figref{Fig:NVpower}{a}. Similar effects have been observed for laser wavelength 1064~nm and below \cite{Iakoubovskii2000, Kupriyanov2000, Manson2005, Aslam2013, Lai2013, Geiselmann2013, Neukirch2013}. We note that the 1550 nm laser is better than a 1064 nm laser for the nanodiamond NV centers. More than 70\% of bare nanodiamonds trapped by a 1550 nm laser show a strong fluorescence signal, while only a few percent of bare nanodiamonds (even though each nanodiamond contains about 500 NV centers on average) and 10-20\% silica-coated nanodiamonds trapped by a 1064 nm laser produce photoluminescence \cite{Neukirch2015,Neukirch2013}. The ionization not only reduces the fluorescence strength, but also lessens the ESR contrast. When the recombination of NV$^0$ to NV$^-$ occurs \cite{Aslam2013}, the probability of NV$^0$ ending up in $|m_\mathrm{s}=0\rangle$ state will be 1/3. Thus the ionization due to trapping laser will reduce the overall population of $m_\mathrm{s}=0$  \cite{Fuchs2010} and reduces the ESR contrast.

\vspace{0.5cm}

\textbf{Electron spin resonance in vacuum.}
Since nanodiamond temperature increases in vacuum, the ESR spectrum shifts to the left as shown in \figref{Fig:NVpressure}{a}. From atmospheric pressure to 31~Torr, the diamond temperature can rise from 300~K to above 450~K (\figref{Fig:NVpressure}{b}). When the mean free path of gas molecules is larger than the size of the nanodiamond, the cooling due to the surrounding gas is proportional to the pressure. So the increase of the temperature of the nanodiamond is inversely proportional to the pressure: $\Delta T\propto 1/P_\mathrm{gas}$ (see Methods), as shown by the solid curves in \figref{Fig:NVpressure}{b}.
When the temperature rises in vacuum, one would expect the ESR contrast to decrease. Counter intuitively, the data show that the ESR contrasts of levitated nanodiamonds increase by more than a factor of 2 when the pressure is decreased from atmospheric pressure to 31~Torr (\figref{Fig:NVpressure}{a,c}). This important phenomenon has not been observed  before since the ESR of a levitated nanodiamond  has not been studied in a vacuum environment prior to our work. The latest relevant work  reported the ESR of a levitated nanodiamond in atmospheric air \cite{Neukirch2015}, but not in vacuum.  Using a 1550~nm trapping laser instead of a 1064 nm laser in Ref. \cite{Neukirch2015}, we observe more than $70\%$ of nanodiamonds  in the optical trap are fluorescent, which is crucial for our investigation of ESR in vacuum.

To verify that the increase in ESR contrast is a reversible process instead of a permanent change of the nanodiamond, we performed an experiment in which the chamber was first pumped from atmospheric pressure to 74~Torr, and then brought back to atmospheric pressure. The contrast at the later atmospheric pressure is slightly higher than the initial atmospheric pressure (red square markers in \figref{Fig:NVpressure}{c}), but significantly smaller than the ESR contrast at low pressure. The slight increase of the ESR contrast at the later atmospheric pressure can be explained by the purification of the nanodiamond after exposed to vacuum \cite{Xu2002, Wolcott2014}. However, the main change of the ESR contrast is reversible.
If the nanodiamond is permanently changed, one would expect the nanodiamond to maintain a high ESR contrast when the pressure is brought back to atmospheric level from 74~Torr, which is opposite to the measured data. So there is no significant change of the nanodiamond after exposed to low vacuum. Therefore, the reduction of surrounding and absorbed air molecules is the primary source of the ESR enhancement.

\begin{figure}[t]
\begin{minipage}{3.5in}
\includegraphics[scale=0.9]{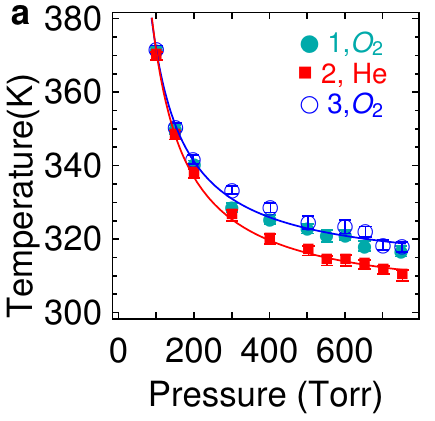}
\includegraphics[scale=0.9]{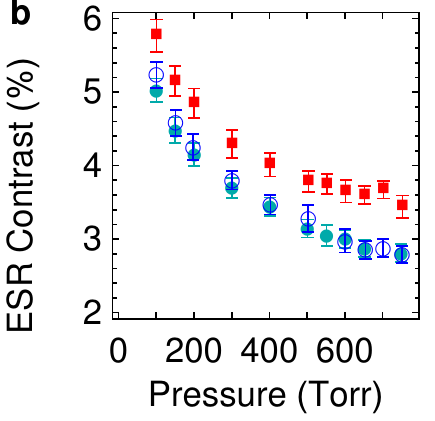} \\
\includegraphics[scale=0.9]{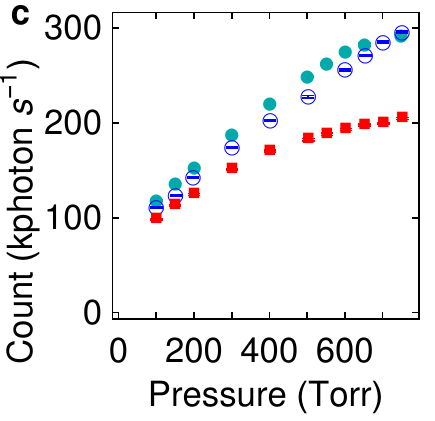}
\includegraphics[scale=0.9]{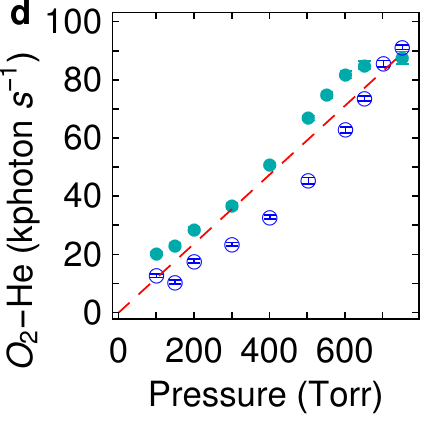}
\end{minipage}
\caption{\textbf{Electron spin resonance in different gases.}
\textbf{a}, The measured temperature of a nanodiamond as a function of the oxygen or helium pressure. From 750 Torr to low vacuum, its temperature changes from 300~K to 380~K and beyond.  The blue (cyan) circle markers represent the data set No. 1 (No. 3) which is taken in oxygen, and the red square markers represent the data set No. 2 which is taken in helium (concentration $>99\%$).
\textbf{b}, The ESR contrast  increases when the gas pressure decreases. The  ESR contrast in helium is higher than the one in oxygen.
\textbf{c}, The total fluorescence count as a function of pressure. The fluorescence signal in helium is lower than the one in oxygen.
\textbf{d},  The count difference between oxygen and helium as a function of pressure. Red dashed line represents the linear fit of the data.
The powers of the 1550~nm trapping laser and the $532$~nm excitation laser are  always held constant with a peak intensity of about 200~($\mathrm{\mu W~\mu m^{-2}}$). 
The error bars of temperature measurements and ESR contrast in \textbf{a,b} are obtained from the standard errors of the fitted parameters from the ESR spectra data. 
The error bars of photon counts  in \textbf{c,d} are the standard deviations of about 70 measurements of total fluorescence signal from the nanodiamond. Each measurement takes 5.3 seconds. 
}
\label{Fig:NVgases}
\end{figure}

\textbf{ESR in different gases.}
To further understand the effects of the surrounding gas on levitated nanodiamond NV centers, we changed the surrounding gas between oxygen and helium repeatedly while a nanodiamond was levitated continuously for many hours.  As shown in \figref{Fig:NVgases}, we first took the data set No. 1 in oxygen. We then changed the gas in the chamber to helium and took the data set No. 2. At last, we changed the gas back to oxygen and took the data set No. 3.  The filled helium or oxygen gas in the chamber has a purity larger than $99\%$, achieved by repeatedly evacuating and filling the chamber several times with a desired gas. The data set No. 1 and data set No. 3 are essentially the same (\figref{Fig:NVgases}), showing the observed effects are reversible and the temporal drift of our system in several hours is negligible.

Our results demonstrate that oxygen and helium gases have different effects on both the ESR contrast and the fluorescence strength of levitated nanodiamond NV centers (\figref{Fig:NVgases}). The ESR contrast of nanodiamond NV centers in helium  is about 25\% higher than that in oxygen (\figref{Fig:NVgases}{b}), while the fluorescence strength in helium is about 30\% lower than that in oxygen near atmospheric pressure(\figref{Fig:NVgases}{c}). When the pressure decreases, the ESR contrast increases while the fluorescence signal decreases in both helium and oxygen.
The different effects of oxygen and helium on the ESR contrast and fluorescence cannot be explained by their temperature difference, e.g., the fluorescence strength in helium is lower than that in oxygen even through the temperature in helium is lower (the thermal conductivity of helium is higher). In general, the fluorescence strength should increase when the temperature decreases (\figref{Fig:NVgases}{a}, {c}) \cite{Plakhotnik2010, Toyli2012}.

The observed phenomena in \figref{Fig:NVgases} can be explained by the reduction of low-quality negatively-charged NV$^-$ centers near the surface due to the reduction of oxygen surface termination \cite{FuAPL2010, HaufPRB2011, Grotz2012, OhashiNL2013}, and a moderate increase of the temperature that quenches low-quality surface NV$^-$ centers \cite{Plakhotnik2010} without significantly affecting high-quality NV$^-$ centers at the center of the nanodiamond\cite{Toyli2012}.
Comparing to NV centers far away from the surface \cite{Toyli2012}, NV centers near the surface have stronger non-radioactive decays, and thus a lower ESR contrast and stronger temperature dependence \cite{Plakhotnik2010}. It has been observed that the fluorescence of NV centers in 20-30 nm-diameter nanodiamonds decreased by 10--40\% when the temperature increased from 300 K to 400 K \cite{Plakhotnik2010}, accompanying by a decrease of the lifetime of the excited state due to non-radioactive decay. In contrast, the fluorescence of a NV center inside a bulk diamond only decreases by a few percents when the temperature is increased from 300 K to 450 K \cite{Toyli2012}. In our experiment, the trapped nanodiamond has about 500 NV centers and a diameter of about 100 nm. When the pressure decreases,  the fluorescence of low-quality surface NV centers are suppressed at an increased temperature (\figref{Fig:NVgases}{c}). To roughly estimate the thickness of the surface shell where the fluorescence of NV centers is quenched, we assume that NV centers are uniformly distributed in the nanodiamond and the fluorescence signal $C_\mathrm{inner}$ of a levitated nanodiamond in low vacuum comes from NV centers within the inner core of radius $r_\mathrm{inner}$.  The $r_\mathrm{inner}$ can be estimated by comparing $C_\mathrm{inner}$ to the fluorescence count $C_\mathrm{atm}$ at atmospheric pressure: $C_\mathrm{inner}/C_\mathrm{atm} \approx r_\mathrm{inner}^3/r^3$, where $r$ is the radius of the nanodiamond. So $r-r_\mathrm{inner}$ is approximately the thickness of the surface shell.  The thickness of the surface shell is estimated to be about 10$-$15 nm for the levitated nanodiamond at 100 Torr  (Fig. 4c).
Because the high-quality NV centers at the center of the 100 nm nanodiamond are largely unaffected by the temperature increase while the low-quality NV centers near the surface are quenched, the overall ESR contrast  increases due to the average effect of 500 NV centers (high-quality and low-quality) as shown in \figref{Fig:NVgases}{b}.

Moreover, a large fraction of NV centers in a nanodiamond are in NV$^0$ charge state \cite{Vertical2009,Grotz2012}, which have low fluorescence signal and no ESR near 2.8 GHz. The oxygen termination allows more NV centers in NV$^-$ charge state. Thus the fluorescence strength increases when the levitate nanodiamond is surrounded by oxygen (\figref{Fig:NVgases}{c}). However, the NV$^-$ centers enabled by oxygen termination are near the surface and have low contrast. So the ESR contrast decreases in oxygen, as shown in \figref{Fig:NVgases}{b}.

While oxygen has been used as permanent surface termination \cite{FuAPL2010,HaufPRB2011,Grotz2012,OhashiNL2013}, here we report that  this  can also happen in air near room temperature and is reversible. Because the effects are reversible, nanodiamond NV centers can be used for oxygen gas sensing repeatedly.
 Using the fluorescence signal in helium gases as the background correction for the thermal effect, the count difference in oxygen and helium gases  exhibits roughly linear dependence on the pressure. Although the total fluorescence counts (\figref{Fig:NVgases}{c}) show a nonlinear scaling, the difference between the counts in oxygen and helium gases (\figref{Fig:NVgases}{d}) trends to be proportional to the pressure in the range of our data. As a first order approximation, we fit the data with a line (dashed line in \figref{Fig:NVgases}{d}). We found our current un-optimized imaging setup can detect about 100~$
\mathrm{ photon ~Torr^{-1} ~ s^{-1}}$. When the oxygen pressure is very large, we expect the signal to deviate from the linear fit because the maximum signal will be limited by the number of NV centers in the nanodiamond.
For future applications in oxygen gas sensing, we can simply put nanodiamonds on a substrate instead of using optical levitation.  Oxygen gas sensors are extensively used to monitor the oxygen concentration in exhaust gases of internal combustion engines in automobiles, and in medical instruments such as anesthesia monitors and respirators. The most common oxygen gas sensor is the zirconia sensor \cite{Fleming1977,Brailsford1997},  which is an electrochemical fuel cell that consumes some oxygen to generate a voltage output depending on the concentration of the oxygen, and only works effectively after being heated up. Comparing to zirconia oxygen sensors, the nanodiamond oxygen sensors do not consume oxygen and do not need to be heated up in order to work. Nanodiamond sensors will also be much smaller. Multiple nanodiamonds can be used to improve the sensitivity. An area of $5~\mu m\times 5~\mu m$ can have thousands of nanodiamonds.

\section{Discussion}
In this paper, we optically levitate a nanodiamond  and demonstrate electron spin control of its built-in NV centers in low vacuum and different gases.
We observed that the ESR contrast of NV centers increases when the air pressure decreases.
We also observe that oxygen and helium gases have different effects on both the photoluminescence and the ESR contrast of nanodiamond NV centers. While more detailed studies are required to fully understand this phenomenon, our observation suggests a potential application of nanodiamond NV centers for  oxygen gas sensing. The  increase of ESR contrast in low vacuum  can improve the sensitivity of other NV center devices such as nanodiamond magnetic/electric field sensors where the NV centers are near the diamond surfaces. We also study the effects of trapping laser and measure the absolute internal temperature of the levitated nanodiamond in vacuum to better understand this system. The results show that the internal temperature of an optically levitated nanodiamond increases significantly in low vacuum. This is because the commercial nanodiamonds used in this work are not pure and have large optical absorption. Better nanodiamond samples with negligible absorption \cite{Andrich2014} at the trapping wavelength  are required  to be optically trapped in high vacuum.

\section{Methods}
\textbf{Experimental setup}.
A mixture of nanodiamonds  (Adamas ND-NV-100nm-COOH) and water at density  about 30~$\mu$g/ml is launched into a vacuum chamber using a Mabis ultrasonic nebulizer for optical trapping.
A nanodiamond is captured and trapped by a 1550~nm laser beam at the focus of a  NA=0.85 infrared objective lens  (\figref{Fig:Apparatus}{a}). The position detector monitors \cite{Li2010} the center of mass motion of the nanodiamond using the  trapping beam after exiting the vacuum chamber. The NV centers are optically excited by a 532 nm laser beam guided by a beam splitter and a 950 nm long pass dichroic mirror. The fluorescence signal of the NV centers are detected by an Andor spectrometer and a Newton EMCCD (electron multiplying charge coupled device) camera. The electron spin of NV centers are controlled using a microwave antenna \cite{Horowitz2012} at a distance of 0.5 mm from the nanodiamond. The gas pressure inside the vacuum chamber is measured by an absolute piezo sensor for pressures above 10 Torr, and a micropirani sensor for lower pressures.

\begin{figure*}[t]
	\begin{minipage}{6.6in}
		\includegraphics[scale=0.9]{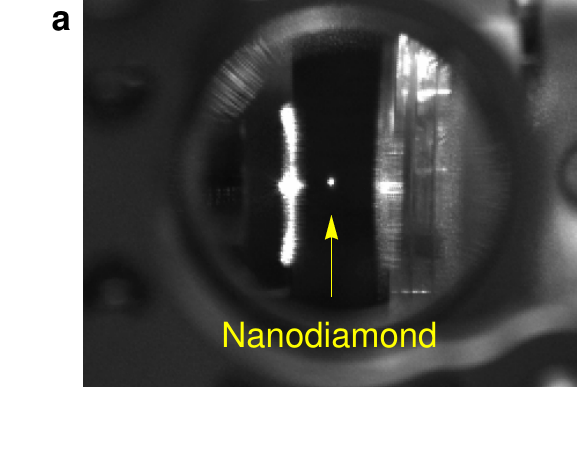}
		\includegraphics[scale=0.9]{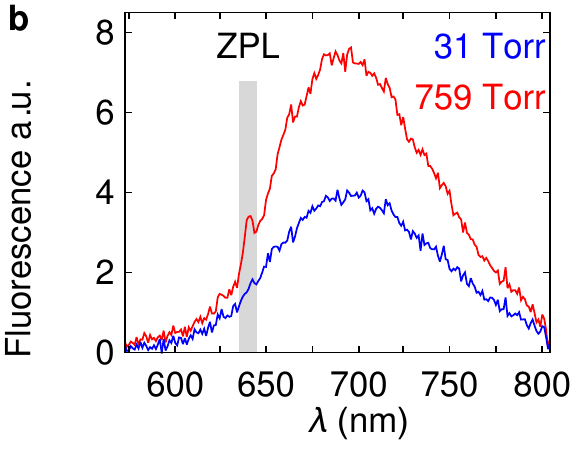}
		\includegraphics[scale=0.9]{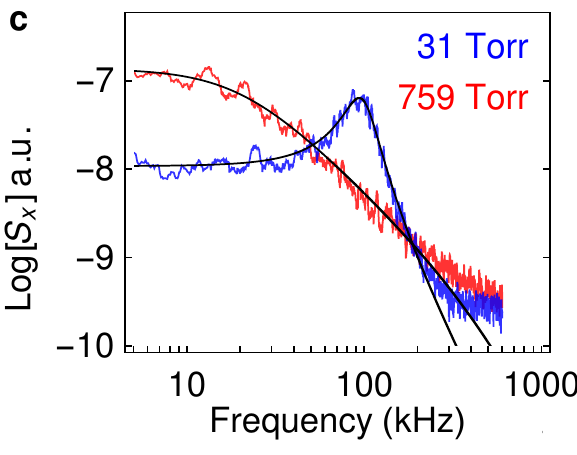}	\\	
		\includegraphics[scale=0.9]{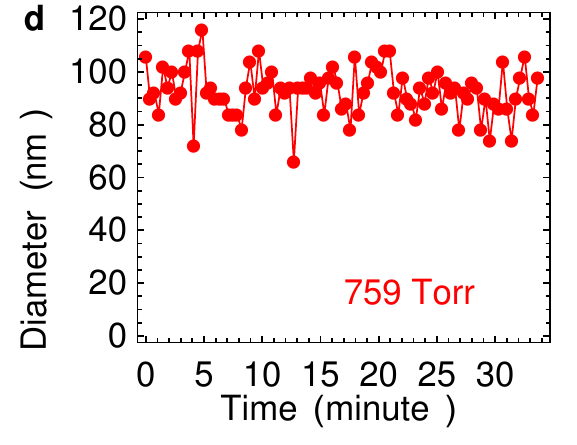}
		\includegraphics[scale=0.9]{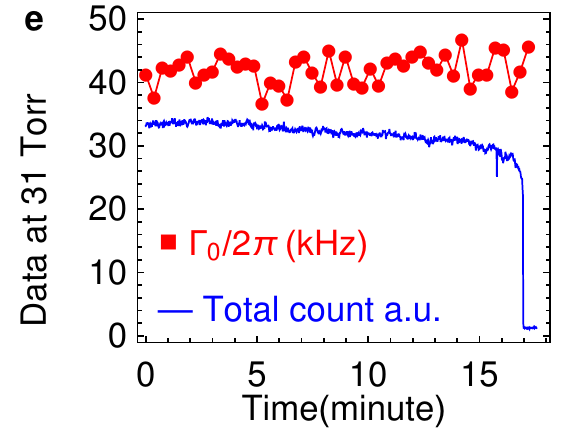}	
		\includegraphics[scale=0.9]{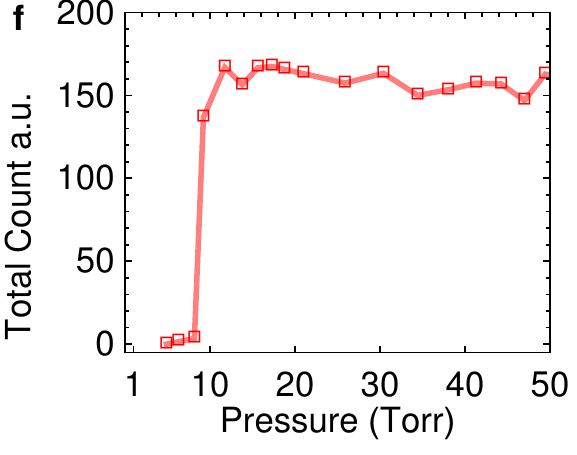}	
	\end{minipage}
	\caption{\textbf{A levitated nanodiamond in low vacuum.} \textbf{a}, An image of an optically levitated nanodiamond (bright white spot) inside the vacuum chamber.
	\textbf{b}, Typical fluorescence spectra  at atmospheric pressure (759 Torr) and in low vacuum (31~Torr).  The nanodiamond is excited with a $30~\mu W$ green laser. The zero phonon line of NV$^-$ is visible near 640 nm.
	\textbf{c}, Power spectral density of the center-of-mass motion of a nanodiamond trapped at two different pressures (red: 759 Torr; blue: 31 Torr). A resonant peak appears in low vacuum. The solid black curves are theoretical fits (see \eqnref{Eqn:powerspectra}).
	\textbf{d}, The hydrodynamic diameter of a trapped nanodiamond calculated from the measured viscosity damping $\Gamma_0$ (see \eqnref{Eqn:Gamma0}).
	\textbf{e}, The viscous damping factor $\Gamma_0$ and total fluorescence count of a levitated nanodiamond at 31~Torr. The nanodiamond is lost at the right side of this figure. The $\Gamma_0$ is fairly constant, suggesting there is no change in the particle size before its loss.  The same nanodiamond is used in figures \textbf{b}, \textbf{c}, \textbf{d} and \textbf{e}.
	\textbf{f}, Total fluorescence count of an optically trapped nanodiamond when the chamber is pumped to vacuum. The nanodiamond is lost at 9~Torr. a.u. denotes an arbitrary unit.
	}
\label{Fig:Spectrum}
\end{figure*}

\textbf{Optical levitation of a nanodiamond in low vacuum.}
To verify the trapped nanodiamond  (\figref{Fig:Spectrum}{a}) has NV centers, we excite the particle with a 532 nm green laser beam and analyze the fluorescence signal using a spectrometer with an EMCCD (\figref{Fig:Apparatus}{a}). One unique signature of NV$^-$ center is the zero phonon line (ZPL) around 637~nm due to the 1.94~eV  transition between the $^3 A_2$ ground state and the $^3E$ excited state as depicted in \figref{Fig:Apparatus}{b} \cite{Acosta2010, Aslam2013}. At room temperature, the visible fluorescence spectrum covers from $600-800$~nm due to thermal phonon broadening. In \figref{Fig:Spectrum}{b},  the fluorescence spectrum indicates a ZPL at $ 640$~nm. This small shift of ZPL from 637 nm is because of temperature effect \cite{Chen2011,Davies1976,Gruber1997}.
We observe that the  visible fluorescent strength of levitated NV centers  decreases and the ZPL becomes weaker in vacuum (\figref{Fig:Spectrum}{b}). When the air pressure decreases, the nanodiamond temperature rises because the cooling rate from air molecules reduces while the heating from the trapping laser remains the same.

Besides optical properties of NV$^-$ centers, we also monitor the center-of-mass   motion (CMM) of a trapped nanodiamond  using the $x$-axis position detector \cite{Li2010}.
At atmospheric pressure, Brownian motion due to collisions with air molecules predominates the harmonics oscillation due to the trapping potential. In low vacuum, the harmonic motion dominates. As a result, a resonant peak ($\approx$ 100~kHz) appears in the spectrum of power spectral density (PSD) of the CMM at low pressure (\figref{Fig:Spectrum}{c}). Fitting the PSD data to theory (see \eqnref{Eqn:powerspectra}) reveals useful  information about the optomechanics system \cite{Aspelmeyer2014}, including trapping frequency $\Omega_0$ and viscous damping factor due to air molecules $\Gamma_0$. 
The trapping frequency $\Omega_0/2\pi$ is about 100~kHz and viscosity damping coefficient $\Gamma_0/2\pi$ goes from about $500$ kHz at atmospheric pressure to about 40~kHz at 31~Torr.
From the fitting parameter $\Gamma_0/2\pi$, the hydrodynamic diameter of nanodiamond, $2 r=94 \pm 7$~nm (see \eqnref{Eqn:Gamma0}), remains constant for over 30 minutes (\figref{Fig:Spectrum}{d}). This result is consistent with the manufacture specification size of $100$~nm. In fact, we observed that a single nanodiamond remains in the trap for more than 200 hours at atmospheric pressure and in low vacuum. A trapped nanodiamond will be lost, however, when the air pressure is below a critical value. \figref{Fig:Spectrum}{f} shows a nanodiamond is lost when the pressure is reduced to 9~Torr. We also observed some nanodiamonds trapped at slightly lower pressures.

\textbf{Particle loss mechanism.}
Combining the  fluorescence signal and center-of-mass motion together provides an insight into the loss mechanism of levitated nanodiamonds in vacuum. As shown in \figref{Fig:Spectrum}{e}, a levitated nanodiamond is lost after being held at 31~Torr for 17 minutes. The total count of the fluorescence signal first decreases gradually and then drops rapidly within the last 60 seconds before the nanodiamond is lost (\figref{Fig:Spectrum}{e}). Based on the fluorescence signal alone, one might conclude that the nanodiamond is gradually burned in residual air in the vacuum chamber. However, the viscous damping factor $\Gamma_0/2\pi$ which is related to the particle size (see \eqnref{Eqn:Gamma0}) is fairly constant at 31~Torr. So the size of the nanodiamond does not change significantly before it is lost.  This is consistent with the  internal temperature of the nanodiamond measured by the ESR which will be discussed later. For the nanodiamond shown in (\figref{Fig:Spectrum}{e}), its internal temperature is about 450~K (177 $^o$C). According to a former investigation, the size reduction speed of nanodiamond due to  air oxidization at atmospheric pressure is about 1 nm/hour at 770 K.
At 450~K, the size reduction of a nanodiamond  should be negligible, especially because the oxygen pressure is low in low vacuum.  So the loss of a trapped nanodiamond is unlikely due to oxidization, instead it might relate to the Brownian motion due to the rising temperature of the nanodiamond.
When the temperature of the nanodiamond increases, the effective temperature of surrounding gases increases \cite{Millen2014}. Since the mean squared displacement of the Brownian motion, $\langle x^2 \rangle$, is proportional to effective temperature of surrounding gases, the particle will explore a larger volume and be more likely to escape from the trapping region.
The loss of levitated nanodiamond in an optically trap is similar to Kramer's escape rate problem describing the escape rate of a particle from a potential barrier, $R_\mathrm{escape} \propto \frac{1}{\Gamma_0} e^{-E_\mathrm{b}/{(k_\mathrm{B} T)}} $ \cite{kramers1940brownian}, here $\Gamma_0$ is the viscous damping factor,  $E_\mathrm{b}$ is the potential barrier, $k_\mathrm{B}$ is Boltzmann's constant, and $T$ is the nanodiamond temperature. The escape rate increases with increasing temperature of the particle.
A larger vibration amplitude can lead to a decrease of the fluorescence signal when nanodiamond randomly walks away the trapping region (overlapping with the focus of the visible imaging system).
 Though, more studies are required to further clarify the loss mechanism.

\textbf{Center-of-mass motion calibration}.
When the nanodiamond is at the thermal equilibrium, the power spectra density (PSD) of nanodiamond CMM is given by \cite{Berg04}
\begin{eqnarray}
	S_x(\omega) = S_0 \frac{\Gamma_0}{(\Omega_x^2 - \omega^2)^2 - \omega^2 \Gamma_0^2}
	\label{Eqn:powerspectra}
\end{eqnarray}
here $S_0 = 2k_\mathrm{B} T/m$, $T$ is the air temperature, $m$ is the nanodiamond mass, $\Gamma_0$ is the viscous damping factor due to the air, $\Omega_x$ is the trapping frequency in the $x$-axis, and $\omega$ is the observation frequency.
Fitting the CMM power spectral density data in \figref{Fig:Spectrum}{c} to \eqnref{Eqn:powerspectra}, one can extract three fitting parameters $S_0, \Gamma_0$, and $\Omega_x$.
In general, the fit works best when the $\omega_x$ is comparable to or larger than $\Gamma_0$ (i.e. when pressure is below 300 Torr). The trapping frequency depends on the laser trapping power. Since the refractive index of air is 1.0027, similar to vacuum, the trapping potential in air should be essentially the same as in vacuum. The trapping frequency $\Omega_x$ does not change much from 300 Torr to 10 Torr according to the data. Therefore, the trapping frequency is presumably constant.

\textbf{Particle size calculation.}
Since it is difficult (and not important)  to know the exact shape and size of the nanodiamond (which can be irregular), we estimate its hydrodynamic size from the viscous damping factor as \cite{Beresnev1990, Li2011}
\begin{eqnarray}
	\Gamma_0 = \frac{6\pi \eta r}{\rho (4/3 \pi r^3)} \frac{0.619}{0.619 + \mathrm{Kn}}(1+c_\mathrm{K})
	\label{Eqn:Gamma0}
\end{eqnarray}
where $\eta$ is the dynamic viscosity coefficient of the air, $r$ is the hydrodynamic radius of the nanodiamond, $\mathrm{Kn}=s/r$ is the Knudsen number with $s$ being the mean the free path of air molecules, and $c_\mathrm{K} = 0.31 \mathrm{Kn}/(0.785 + 1.152 \mathrm{Kn} + \mathrm{Kn}^2)$.  The mean free path is linearly dependent on the viscosity as \cite{Beresnev1990}
\begin{eqnarray}
	s =  \frac{\eta}{P} \sqrt{\frac{\pi k_\mathrm{B} T}{2m_\mathrm{air}}}
	\label{Eqn:freepath}
\end{eqnarray}
here $P$ is the pressure and $m_\mathrm{air}$ is the mass of air molecule. In 1866, Maxwell demonstrated that the viscosity of air maintains constant from atmospheric pressure down to a few Torr \cite{Maxwell1867}.  At atmospheric pressure, the nanodiamond in \figref{Fig:Spectrum}{d} ($\approx 300$~K) is presumably in thermal equilibrium with air molecules. The temperature dependence of the viscosity can be calculated using Sutherland's formula \cite{Sutherland1983} which yields $\eta\approx 18.52 \times 10^{-6}$ (Pa s) and $s\approx 67$~nm at room temperature (296~K). The the viscous damping factor is $\Gamma_0 \approx 500$~kHz obtained from the power spectral density fit of \figref{Fig:Spectrum}{c} with \eqnref{Eqn:powerspectra}. Solving for $r$ in \eqnref{Eqn:Gamma0}, we obtain the nanodiamond diameter of $94 \pm 7$~nm while monitoring the particle size for over 30 minutes (\figref{Fig:Spectrum}{d}). At low vacuum, the nanodiamond is not in thermal equilibrium with the surrounding gas due to its high internal temperature, \eqnref{Eqn:Gamma0} is not suitable to calculate the particle size.

\textbf{ESR Technique}. The ESR experiments were conducted using an optically detected magnetic resonance (ODMR) technique \cite{Van1988,Gruber1997}. We employed a similar microwave antenna design and the microwave lock-in technique described in Refs. \cite{Horowitz2012}. To obtain the ESR signal, the electron spin states are excited $|m_\mathrm{s}=0\rangle \leftrightarrow |m_\mathrm{s}=\pm 1\rangle$ by microwave pulses.
For a given frequency, the antenna delivers a sequence of 500~ms alternative on/off microwave pulses to control the electron spin of NV centers. When the microwave pulse is on, the NV ground spin state $|m_\mathrm{s}=0\rangle$ is excited to $|m_\mathrm{s}=\pm 1\rangle$, and vice versa. At the same time, the visible fluorescence spectrum of the diamond is acquired using the spectrometer (\figref{Fig:Apparatus}{a}).
Since the visible fluorescence signal of $|m_\mathrm{s}=\pm 1\rangle$ is weaker than the $|m_\mathrm{s}=0\rangle$ state, there is a dip in the plot of the normalized $I_\mathrm{PL}$ when ESR happens. The normalized fluorescence signal $I_\mathrm{PL}$  is the ratio of total fluorescence counts with and without microwave excitation, where the resonance peaks occur at the frequencies of minimum $I_\mathrm{PL}$.

\textbf{Nanodiamond temperature.}
The zero-field Hamiltonian of NV center is written as \cite{Balasubramanian2008}
\begin{eqnarray}
	\mathcal{H} = D S^2_z + E(S^2_x - S^2_y)
\end{eqnarray}
Here $D$ is the energy splitting between spin ground state $|m_\mathrm{s}=0\rangle$ and $|m_\mathrm{s} = \pm 1\rangle$, and $E$ is the splitting between $|m_\mathrm{s} = \pm 1\rangle$ due to the broken axial symmetry of the NV center caused by the strain effect \cite{Gruber1997, Lai2009}. This Hamiltonian results in a double peak resonance spectrum as seen in \figref{Fig:NVpower}{b} where $D$ is center of the double peaks, and $E$ is the separation between two resonance peaks. The parameter $E$ is different for each unique NV center. Overall we observe a separation $2E\sim10$ ~ MHz which is consistent with early measurements in bulk diamond \cite{Gruber1997, Acosta2010}. Since the ambient magnetic field was measured to be 600~mG using a commercial magnetometer, the separation between two resonance peaks is mainly contributed by the zero-field parameter $E$. The zero-field splitting parameter $D$ depends on temperature as \cite{Toyli2012}
\begin{eqnarray}
	D(T) = a_0 + a_1 T + a_2 T^2 + a_3 T^3  \nonumber\\
	+ \Delta_\mathrm{pressure} + \Delta_\mathrm{strain}
	\label{Eqn:DT}
\end{eqnarray}
here $a_0=2.8697$~GHz, $a_1=9.7\times 10^{-5}$~GHz/K, $a_2=-3.7\times 10^{-7}$~GHz/K$^2$, $a_3=1.7\times 10^{-10}$~GHz/K$^3$,  $\Delta_\mathrm{pressure} = 1.5$~kHz/bar is the pressure dependence shift \cite{Doherty2014}, and $\Delta_\mathrm{strain}$ is the internal strain effect depending on the particle.

 From atmospheric pressure to 1~Torr, the shift due to pressure is about 1.5~kHz, equivalent to changing nanodiamond temperature by 20~mK, which is negligible compared to the thermal shift. Each individual nanodiamond has a different internal strain $\Delta_{strain}$ about a few MHz \cite{Geiselmann2013}. This could lead to an uncertainty of tens of kelvins in measuring the absolute temperature if $\Delta_\mathrm{strain}$ is neglected.
 As shown in \figref{Fig:NVpower}{c}, the temperature of each nanodiamond increases linearly with the trapping power (within the range of measured data), although of the slope of each line is different. We can use this linear dependence relation to determine the internal strain of each  trapped nanodiamond. One can solve for temperatures $T$ from  $D(T)$  in \eqnref{Eqn:DT} using the parameters $D$ obtained from the double Gaussian fit of ESR spectra \citep{Acosta2010,Horowitz2012}.
Since we know the internal temperature of the nanodiamond is at room temperature (296~K) when there is no trapping laser,   the linear fit in \figref{Fig:NVpower}{c} should intersect the vertical axis at 296~K when the laser power is zero. From this known temperature at the limit of zero trapping power, we determine the internal strain $\Delta_\mathrm{strain}$ of each individual nanodiamond.

 The levitated nanodiamond in a gas or low vacuum is cooled from the collisions with surrounding gas molecules. In the molecular regime, the cooling rate is proportional to the gas pressure. Therefore, the temperature of the nanodiamond depends on the pressure of the surrounding gas approximately as
\begin{eqnarray}
	T = T_0 + \alpha/P_\mathrm{gas}
	\label{Eqn:TempPressure}
\end{eqnarray}
with  $T_0$ being the room temperature, and $\alpha$ being a coefficient that depends the properties of the particle and surrounding gas, and the laser power. This simple equation fits well with our experimental results shown in \figref{Fig:NVpressure}{b} and \figref{Fig:NVgases}{a}.

\textbf{Data Availability} The data that support the findings of this study are available from the corresponding author upon request.

\textbf{Acknowledgements}
We acknowledge the support from  the National Science Foundation under Grant No. 1555035-PHY.
We would like to thank helpful discussions with F. Robicheaux, C. Zu, Z. Yin, N. Zhao,  M. Y. Shalaginov, and V. M. Shalaev.

\textbf{Author Contributions} T.M.H. and T.L.  conceived and designed the project, and analyzed the data.
T.M.H., J.A, J.B. built the experimental apparatus. T.M.H. performed measurements.
T.L. supervised the work. All authors co-wrote the paper.

\textbf{Competing financial interests}
This paper may affect patent application.

\bibliographystyle{nature}

\onecolumngrid
\newpage

\end{document}